\documentclass[final,1p,times]{elsarticle} 
\usepackage{graphicx}
\usepackage{amssymb} 
\usepackage{amsthm} 
\usepackage{lineno}

\newcommand{\sqn}{\mbox{$\sqrt{s_{NN}}$}}

\newcommand{\pp}{\mbox{p+p}}
\newcommand{\pbpb}{\mbox{Pb+Pb}}
\newcommand{\auau}{\mbox{Au+Au}}
\newcommand{\rcp}{\mbox{$R_{\mathrm{CP}}$}}
\newcommand{\raa}{\mbox{$R_{\mathrm{AA}}$}}
\newcommand{\pT}{\mbox{$p_{\mathrm{T}}$}}
\newcommand{\pt}{\mbox{$p_{\mathrm{T}}$}}
\newcommand{\vd}{\mbox{$v_{\mathrm{2}}$}}
\newcommand{\vt}{\mbox{$v_{\mathrm{3}}$}}

\newcommand{\Npart}{\mbox{$\langle N_{\mathrm{part}}\rangle$}}

\newcommand{\vup}{\vspace{-6mm}}

\journal{Nuclear Physics A} 

\begin{document}
\begin{frontmatter} 

\title{High-\pT\ and Jets. A Summary of Results from Quark Matter 2012}

\author[auth1]{Jorge Casalderrey Solana}
\author[auth2]{Alexander Milov}
\address[auth1]{Departament de d'Estructura i Constituents de la Mat\`eria 
and Institut de Ci\`encies del Cosmos (IC-CUB),\\
Universitat de Barcelona, Mart\'i i Franqu\`es 1, 08028 Barcelona, Spain} %%%% check you range and affiliation
\address[auth2]{Department of Physics and Astrophysics, Weizmann Institute of Science, 234 Herzl str. Rehovot, 76100, Israel}

\begin{abstract} 
A broad range of new experimental data and theoretical results on the properties of hadronic matter under extreme conditions have been reported at Quark Matter 2012 conference. At this conference the scientific community was presented with a variety of measurements from the 2011 lead-lead LHC run using  hard probe observables. Many measurements, such as boson-jet correlations, production rates of the b-jets, high precision jet fragmentation and others were shown for the first time. The new data from the LHC was matched by new techniques and analyses coming from RHIC experiments.  This proceeding summarises the new measurements with high-\pT\ particles and jets and attempts to provide a theoretical explanation  for the novel results presented at the conference.
\end{abstract} 

\end{frontmatter} % do not change

\linenumbers % linenumbers are useful for reviewing process

The physics of relativistic Heavy Ion (HI) collisions is one of the most dynamic fields of modern physics. Its success is based on the wealth of experimental data coming from the Relativistic Heavy Ion Collider (RHIC) at Brookhaven National Laboratory and the Large Hadron Collider (LHC) at the European Organisation for Nuclear Research. Studying hard processes is crucial to understanding the properties of the medium created in HI collisions. Experiments at the LHC operate at new energy regime where jets are produced at high rates, which makes them one of the most  interesting observables.
The LHC results are complemented by the analysis of RHIC energy scan data and the possibility to study various colliding species which are crucial to understand the nature of observed effects. Results shown at the Quark Matter 2012 conference include the detailed study of the suppression of high-\pt\ charged particle production and fully reconstructed jets; analysis of jet fragmentation and jet shape modifications in \pbpb\ collisions; the first studies on jets containing b-quarks. Existing results on anisotropy of high-\pt\ particles were extended with the first results of the fully reconstructed jet anisotropy. These measurements provide detailed information about the jet-medium interactions and present new challenges for its theoretical understanding. Jet observables demand an analysis of the behaviour of the whole jet shower in the medium, which go beyond the leading particle energy loss. \\

Modification of measured observables in HI collisions can be expressed in terms of the nuclear modification factor, the ratio of yields measured in the HI system to that measured in the \pp\ (\raa) or peripheral HI system (\rcp), taking into account the geometric factor of nuclei. The \rcp\ of inclusive hadrons measured at different energies is shown in the left panel of Fig.~\ref{fig:raa}\footnote{Here and in the following figures, the layouts of some plots have been modified to satisfy the space constrains. For unmodified plots refer to original publications mentioned in figure captions.}.
\begin{figure}[h!]
\begin{center}
\includegraphics[width=1.0\textwidth]{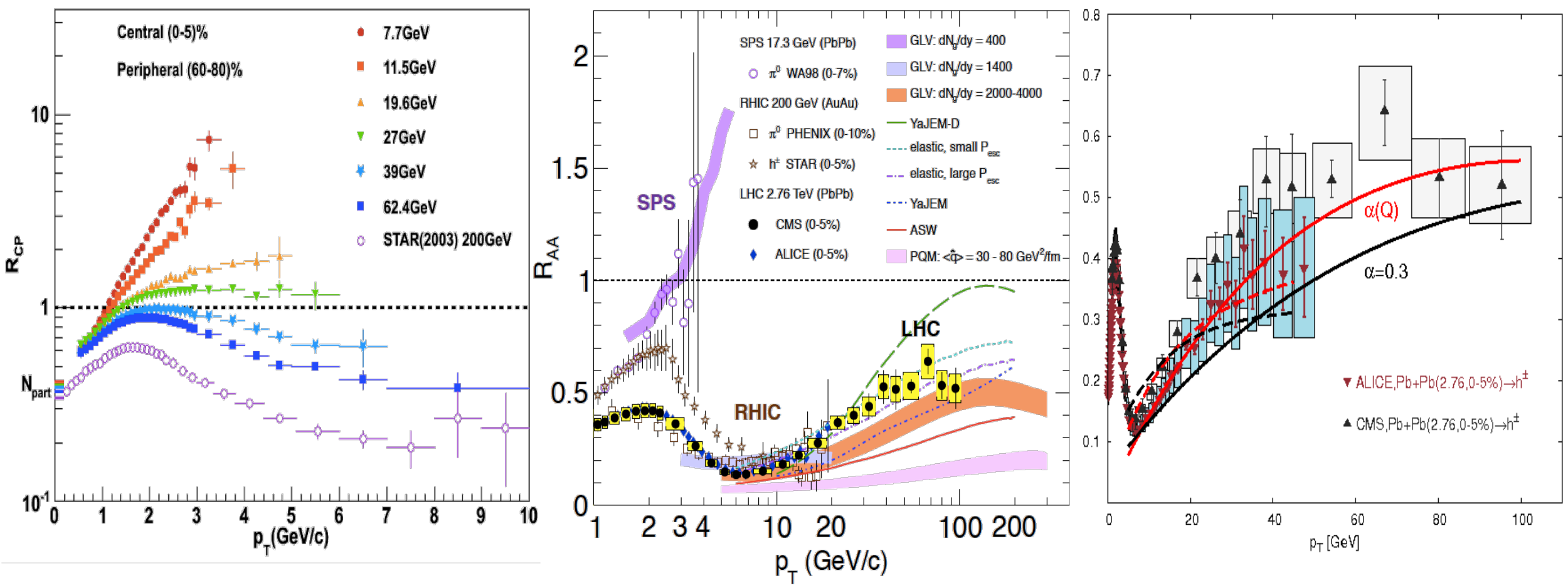}
\end{center}
\vup
\caption{Left: STAR measurement of the \rcp\ for different RHIC energies~\cite{star_rcp}. Middle: The \raa\ measured by CMS~\cite{cms_charged_raa} compared to the top RHIC energy and several models listed in~\cite{cms_charged_raa}. Right: \raa\ compared to the model \cite{buzzatti}.}
\label{fig:raa}
\vspace{-4mm} % CAREFULL !!!!
\end{figure}
The plot shows the onset of suppression at $\sqn\sim30-40$\,GeV that reaches \rcp=0.2 at \pt=7\,GeV at the highest RHIC energy and changes only slightly between RHIC and LHC regimes (middle panel). Since the RHIC spectra are steeper than at the LHC, the same \raa\ would require more energy loss at LHC than at RHIC. At $\pt>80$\,GeV the data from the LHC experiments~\cite{cms_charged_raa,atlas_charged_rcp} indicate that the charged hadron \rcp\ reaches $\sim$0.6 and does not grow with increasing \pt, which puts constrains on the fragmentation functions discussed further in this proceeding.

When comparing the \raa\  data to different models for parton energy loss several conclusions can be drawn. Most of the models which successfully describe the RHIC data correctly predict the general trend of the \pt\ dependence of \raa\ observed at the LHC. However, a quantitative description of higher energy data require some additional tuning. An explicit example of such tuning, shown in the right panel of Fig.~\ref{fig:raa}, was presented in \cite{buzzatti}, pointing out that the large \pt\ range of the LHC results requires taking into account the running of $\alpha_s$ in the energy loss calculations.
 
% A.M.: I suggest a paragraph here
In general, models which can successfully account for the rise of \raa\ at high \pt\ include both radiative (inelastic) processes and collisional (elastic) energy loss but the relative importance of those processes is not the same in different computations. These effects have an impact  on the path length dependence  of \raa;  the simultaneous description of RHIC and the LHC data imposes tight constraints which provide discriminative power among different models~\cite{renkproc}. Difficulties in reproducing the path length dependence or \raa\ led to the exploration of exotic energy and length dependencies of the energy loss mechanism, motivated in part by strong coupling analyses of energy loss via the AdS/CFT correspondence~\cite{Betz:2012qq}.

\begin{figure}[h!]
\begin{center}
\includegraphics[width=1.0\textwidth]{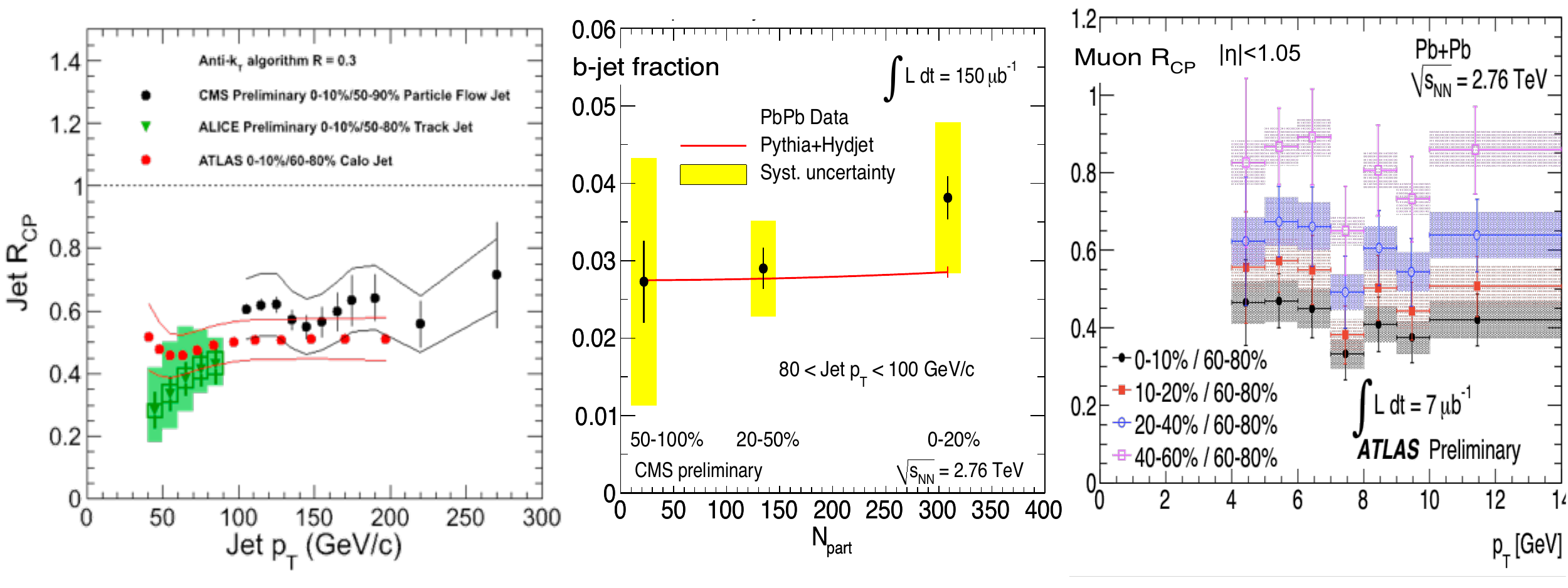}
\end{center}
\vup
\caption{Left: a compilation of the \rcp\ measured for fully reconstructed calorimetric jets by ATLAS~\cite{atlas-jets} (red), CMS~\cite{cms-jets} (black) and track-jets measured by ALICE~\cite{alice-jets} (green). Middle: Fraction of b-jets in different centrality bins measured by CMS~\cite{cms-bjets}. Right: The \rcp\ of muons in jets measured by ATLAS~\cite{atlas-bjets}.}
\label{fig:b_jets}
\vspace{-4mm} % CAREFULL !!!!
\end{figure}

The high-\pt\ particles are predominantly coming from the fragmentation of jets. For the first time, the results on fully corrected jets were presented at the conference by several experiments. The jet \rcp\ measured by the three LHC experiments is shown in the left panel of Fig.~\ref{fig:b_jets}. 
Results from the ATLAS~\cite{atlas-jets} and CMS~\cite{cms-jets} experiments, measuring calorimetric jets, agree within uncertainties. Results from the ALICE experiment~\cite{alice-jets}, measuring track-jets, agree above \pt=70\,GeV with data from ATLAS and CMS. The degree of jet suppression at high-\pt\ is similar to that found with high-\pt\ charged hadrons. The middle panel of the figure shows the fraction of  b-jets~\cite{cms-bjets}. This fraction remains the same within uncertainties for all measured centralities. The right panel of Fig.~\ref{fig:b_jets} presents the measurement~\cite{atlas-bjets} of the \rcp\ of muons associated with jets. The suppression reaches approximately 0.5 in the most central events and is almost independent of \pt. Both results indicate that the b-jets are suppressed with a magnitude similar to that measured for inclusive jets.
\\

\begin{figure}[h!]
\begin{center}
\includegraphics[width=1.0\textwidth]{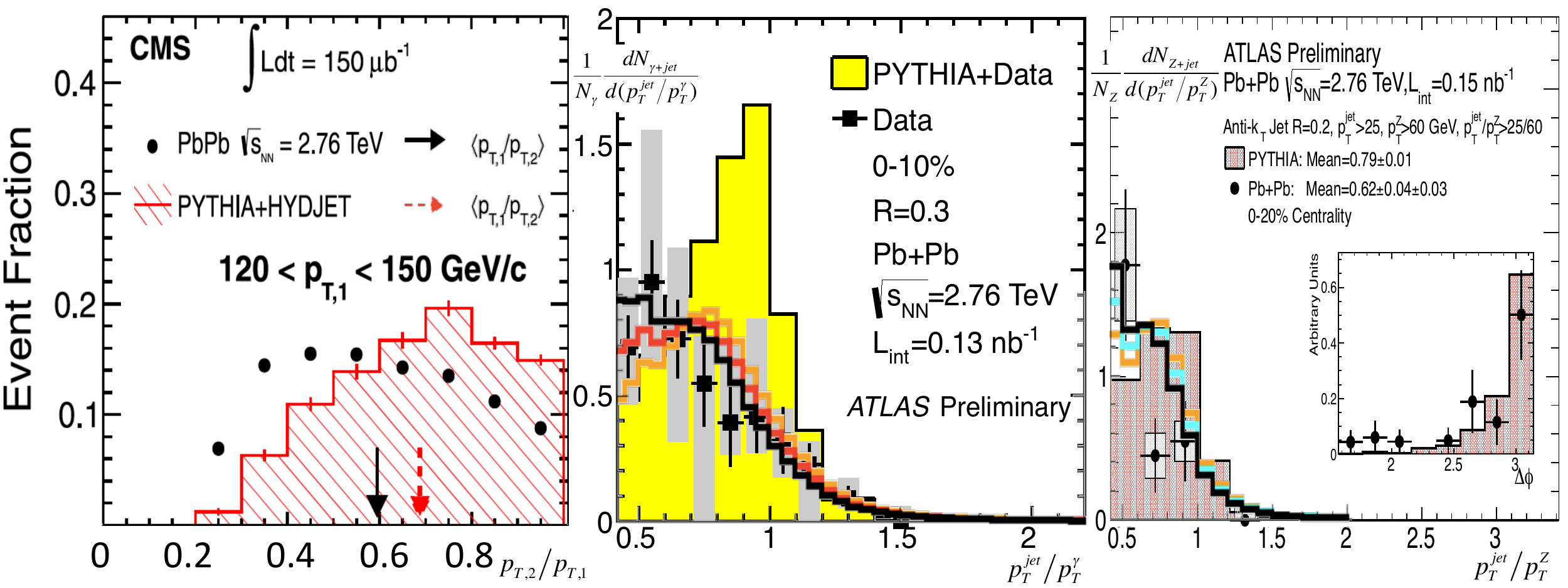}
\end{center}
\vup
\caption{Left: ratio of subleasing to leading jet momentum measured by CMS~\cite{cms-aj2}. Middle and right panels show fully corrected jet to boson momentum ratios measured by ATLAS experiment for photons~\cite{atlas-gjets} and $Z$ bosons~\cite{atlas-zjets} respectively. The curves are based on~\cite{vitev}.}
\label{fig:tag_jets}
\end{figure}

One of the most striking results of the first \pbpb\ run at the LHC is the observation of the di-jet energy asymmetry~\cite{atlas-aj, cms-aj1}. The left panel of Fig.~\ref{fig:tag_jets} shows the momentum balance of two jets reconstructed in opposite hemispheres (black points)~\cite{cms-aj2}. 
The results are compared to a simulation (red histogram) which only takes into account detector effects. A clear reduction of the ratio compared to simulations is seen in the data.

Measuring the momentum balance of two color charged objects gives limited information, since both of them lose energy in the medium. The correlation between jets and bosons provides a better way to quantify jet energy loss by measuring the initial jet momentum using the \pt\ of the boson. %Results on the measurements of jets produced in pairs with isolated photons were presented at the conference by the CMS and ATLAS collaborations~\cite{cms-gjets,atlas-gjets}. 
% show a significant reduction of the jet to boson momentum ratio, which increases with centrality. 
%The first results on the $Z$-jet measurements were also shown by ATLAS~\cite{atlas-zjets} 
Figure~\ref{fig:tag_jets} shows distributions of the ratio of fully corrected jet momentum to the momentum of isolated photon~\cite{atlas-gjets} (middle panel) and $Z$ bosons~\cite{atlas-zjets} (right panel). The plots corresponding to central collisions demonstrate a strong suppression of the data  compared to simulation that does not take into account any energy loss mechanisms. The model prediction shown with lines in the two panels~\cite{vitev} successfully describes the data points.

\begin{figure}[]
\begin{center}
\includegraphics[width=1.0\textwidth]{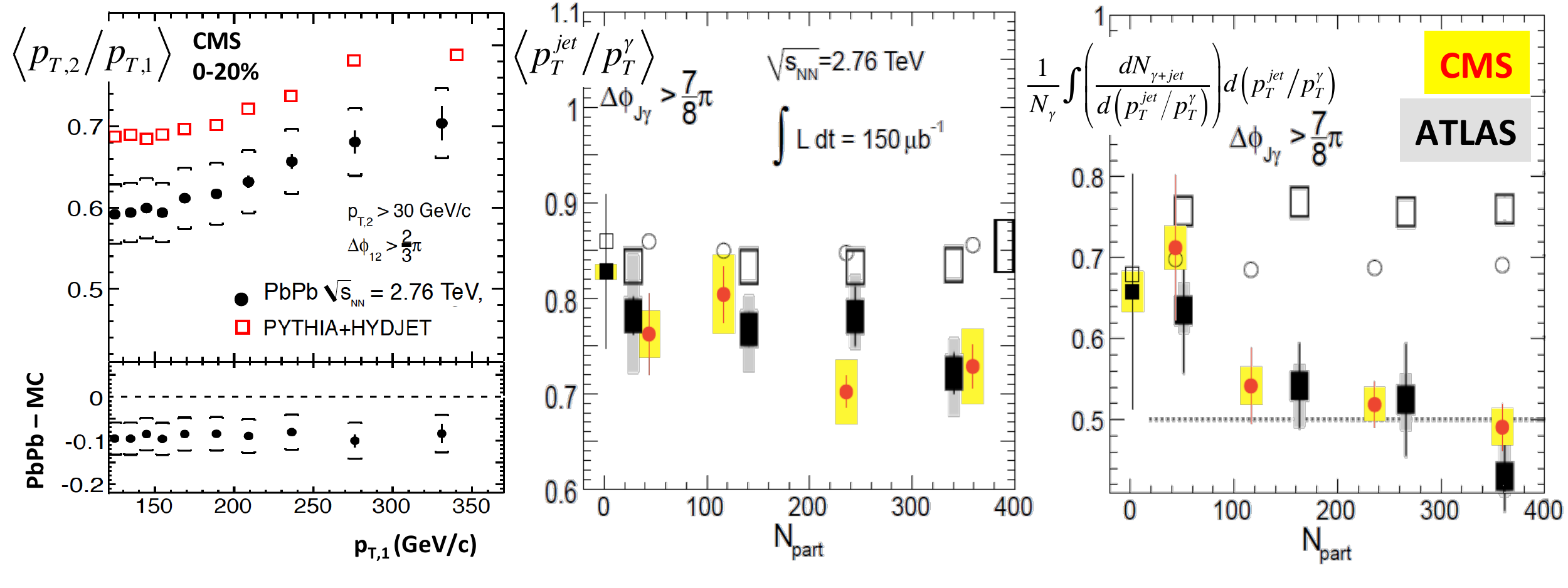}
\end{center}
\vup
\caption{Left: the mean of jet momenta ratio as a function of the leading jet \pt\ measured by  CMS~\cite{cms-aj2}. Middle and right panels show centrality dependence of average jet to photon momentum ratio and the number of boson-jet pairs per boson compiled from  CMS and ATLAS results~\cite{cms-gjets, atlas-gjets}.}
\label{fig:tag_jets_summary} 
\vspace{-4mm} % CAREFULL !!!!
\end{figure}
The left panel of Fig.~\ref{fig:tag_jets_summary} shows the average ratio of sub-leading to leading jet momenta
as a function of the leading jet \pt~\cite{cms-aj2}. For comparison, the open squares in the same plot show the expectation from a simulation which does not include energy loss effects. The data shows significantly stronger imbalance compared to simulation at all measured \pt. The difference between the data and simulation does not depend on the leading jet momentum as shown in the lower part of the left panel. 

To understand the centrality dependence of the  ratios shown in Fig.~\ref{fig:tag_jets}, the middle panel of Fig.~\ref{fig:tag_jets_summary} presents the average ratio of jet and photon momenta as a function of  \Npart\ measured by the CMS~\cite{cms-gjets} and ATLAS~\cite{atlas-gjets} experiments. The results of the two measurements agree.  In peripheral events the experimental data, shown with filled symbols, are consistent with a simulation which takes into account only detector effects (open symbols). At higher centrality the data are significantly lower. The results of the two experiments agree, showing a reduction of the jet-to-boson average momentum ratio with centrality.

Another parameter sensitive to energy loss, shown in the right panel of Fig.~\ref{fig:tag_jets_summary}, is the number of boson-jet pairs per measured boson. Jets in HI can be reliably reconstructed above a certain threshold. Due to the falling jet spectra~\cite{cms-jets,atlas-jets,alice-jets} the fraction of jets which exceed a given threshold depends on the energy loss. The plot shows that  after taking all corrections for the jet reconstruction efficiency and fake rejection in HI collisions this fraction significantly decreases from peripheral to central events.

\begin{figure}[h!]
\begin{center}
\includegraphics[width=1.0\textwidth]{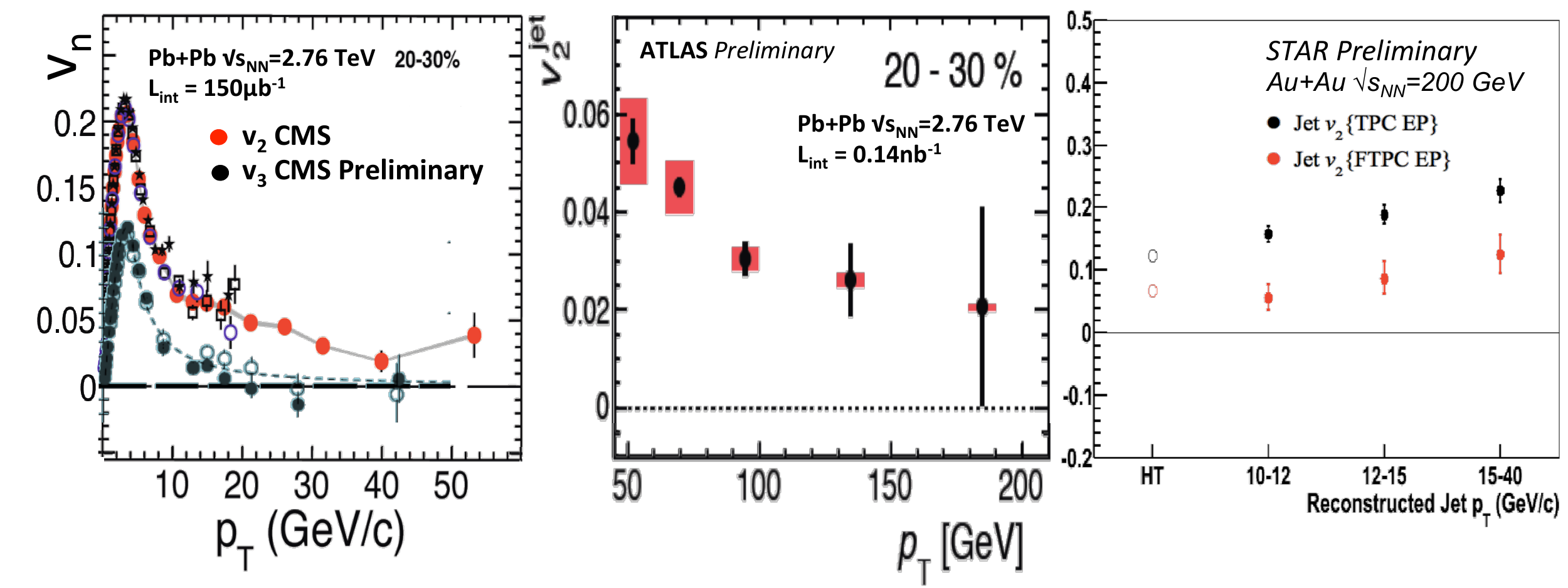}
\end{center}
\vup
\caption{Left: \vd\ (red) and \vt\ (black) harmonic of the charged particle azimuthal distribution measured by CMS~\cite{cms-flow}. Middle: \vd\ measured for fully reconstructed jets measured by ATLAS~\cite{atlas-flow}. Right: jet \vd\ measured by STAR~\cite{star-flow}.}
\label{fig:jets_v2}
\vspace{-4mm} % CAREFULL !!!!
\end{figure}

To quantify the parton energy loss in the medium its path length dependence must be understood. A way to get information on that is to study the azimuthal anisotropy of particle yields. The left panel of Fig.~\ref{fig:jets_v2} shows the second (red) and third (black) harmonics of the charged hadron azimuthal distribution measured up to very high \pt.
The \vt\ originating from fluctuations in the initial geometry of the collision vanishes at $\pt>30$\,GeV, whereas the \vd\ originating from the eccentricity of the system geometry, remains significantly larger than zero at all measured \pt. This anisotropy can be explained by the fact that high-\pt\ particles are more strongly attenuated in the direction of larger transverse dimension of the medium. A quantitatively similar anisotropy is measured at the same collision energy with fully reconstructed jets as shown in the middle panel of Fig.~\ref{fig:jets_v2}. Both results contrast with the \vd\ measured with $Z$ bosons at the LHC~\cite{atlas-z} which is consistent with zero. At RHIC energy, the preliminary results shown in the right panel have much stronger anisotropy, especially at higher pseudo-rapidity~\cite{star-flow}. The \pt\ trend is also different from the measurement at higher \sqn.
%\\% CAREFULL !!!!

The theoretical description of  jet data  poses new challenges since it demands an understanding of the interactions of multiple fragments propagating simultaneously in the medium.  Different calculations~\cite{renkproc,vitevproc,colproc,milhanoproc} which model the data presented above, include both collisional and radiative energy loss mechanisms of secondary shower patrons. Unlike single particle measurements, jet measurements are sensitive to the angular distribution of the lost energy. 
All the models mentioned above include processes with well defined angular structure, either in the form of medium induced radiation~ \cite{vitevproc,colproc, milhanoproc} or in the form of in-medium modification of splitting functions~\cite{renkproc}. Furthermore, a crucial component for describing jet observables is the inclusion of processes without a specific direction of emissions achieved either by a direct transfer of energy to medium particles~\cite{renkproc,vitevproc,colproc} or by the randomization of the softer jet components~\cite{renkproc,milhanoproc}.
Therefore, those processes can be classified according to the angular structure of the lost energy.
%A different classification of those processes can be introduced: on the one hand, all the above models include processes with well defined angular structure, either in the form of medium induced radiation~ \cite{vitevproc,colproc, milhanoproc} or in-medium modification of splitting functions~\cite{renkproc}. On the other, a crucial component for describing di-jet asymmetries is the inclusion of processes without a specific direction of emissions, either by a direct transfer of energy to medium particles~\cite{renkproc,vitevproc,colproc} or by the randomization of the softer jet components~\cite{renkproc,milhanoproc}. 
While the distinction among the processes can only be achieved by a systematic comparison to data, simple kinematic arguments show that the behavior of soft components plays a central role in understanding jet data~\cite{CasalderreySolana:2010eh}. Interesting new advances in describing the soft sector of the jet via a generating functional approach  have been presented in~\cite{fabio}, which provide a theoretical foundation for a probabilistic description of soft jet emissions. \\
 
\begin{figure}[h!]
\begin{center}
\includegraphics[width=1.0\textwidth]{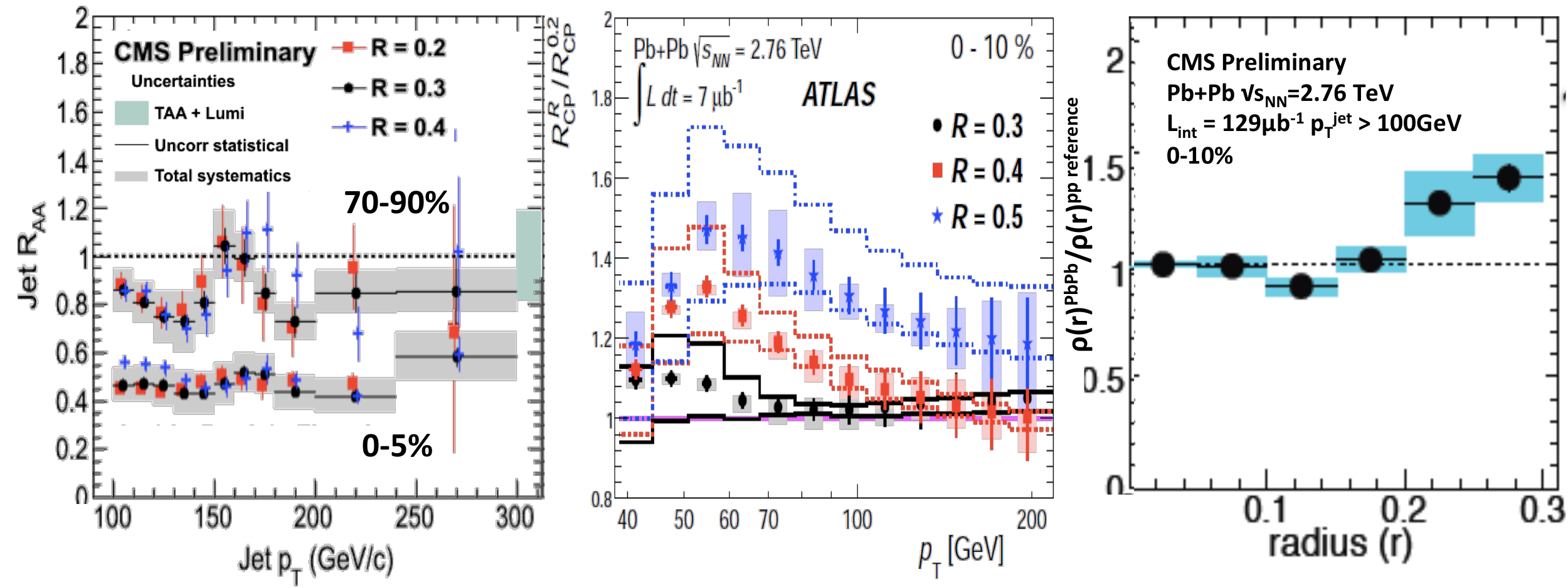}
\end{center}
\vup
\caption{Left: the jet \raa\ measured in two centrality bins for different radius parameter $R$ of reconstructed jets measured by CMS. Middle: the ATLAS result on the double ratio of jet \rcp's measured for jets with different parameter $R$~\cite{atlas-jets}. Right: the ratio of the \pbpb\ and \pp\ jet shapes~\cite{cms-frag} measured by CMS.}
\label{fig:jets_r}
\end{figure}

%%%%%%%%%%%%%%%%%%%%%%%%%%%%%%%%%%%%%%%%%%
 Redistribution of the jet energy is  one of the central questions in understanding parton energy loss. It can be studied by measuring jets with different radius parameter $R$ ~\cite{atlas-jets,cms-frag} as shown in the left panel of Fig.~\ref{fig:jets_r}.
The  magnitude of jet suppression depends on the parameter $R$, which can be better quantified by taking a double ratio of jets $\rcp$ with different $R$ to the \rcp\ with $R=0.2$ as shown in the middle panel of Fig.~\ref{fig:jets_r}. The ratios show lesser suppression of jets with large  $R$ for $\pt<100$\,GeV.

Another way to look at this phenomenon is to study the jet shape as a function of radius. The ratio of jet shape distributions in the most central \pbpb\ collisions and in \pp~\cite{cms-frag} is shown in the right panel of Fig.~\ref{fig:jets_r}. As one can see from the plot there is an enhancement at larger $R$. The results shown in the  figure suggest that the jet energy is redistributed towards larger $R$.

Both the R dependence of \raa\ and the jet shapes provide additional constraints on the interplay between collisional and radiative processes for in-medium jet interactions. As pointed out in \cite{zangproc}, the strong angular dependence of medium induced gluon radiation leads to a significant $R$ dependence of the jet suppression pattern since the emitted jet energy is recovered at sufficiently large R. On the contrary, elastic losses reduce this dependence, since they lead to the randomization of the transported energy by incorporating it into medium particles.  The $R$ dependence observed for the lower jet energies shown  in the middle panel of Fig.~\ref{fig:jets_r} as well as the  "push"  towards larger radii shown in the right panel favours the dominance of radiative processes~\cite{zangproc}. 
However, the same calculations predict that the difference of jet \raa\ with different $R$ remains approximately constant at high \pT, which seems at odds with the measurements in the left panel of Fig.~\ref{fig:jets_r}. Following these arguments, the $R$-independence of \raa\ may point towards a dominance of elastic process at high \pT\, which would be counter intuitive. Additional theoretical studies are needed before drawing firm conclusions. \\
 
\begin{figure}[h!]
\begin{center}
\includegraphics[width=1.0\textwidth]{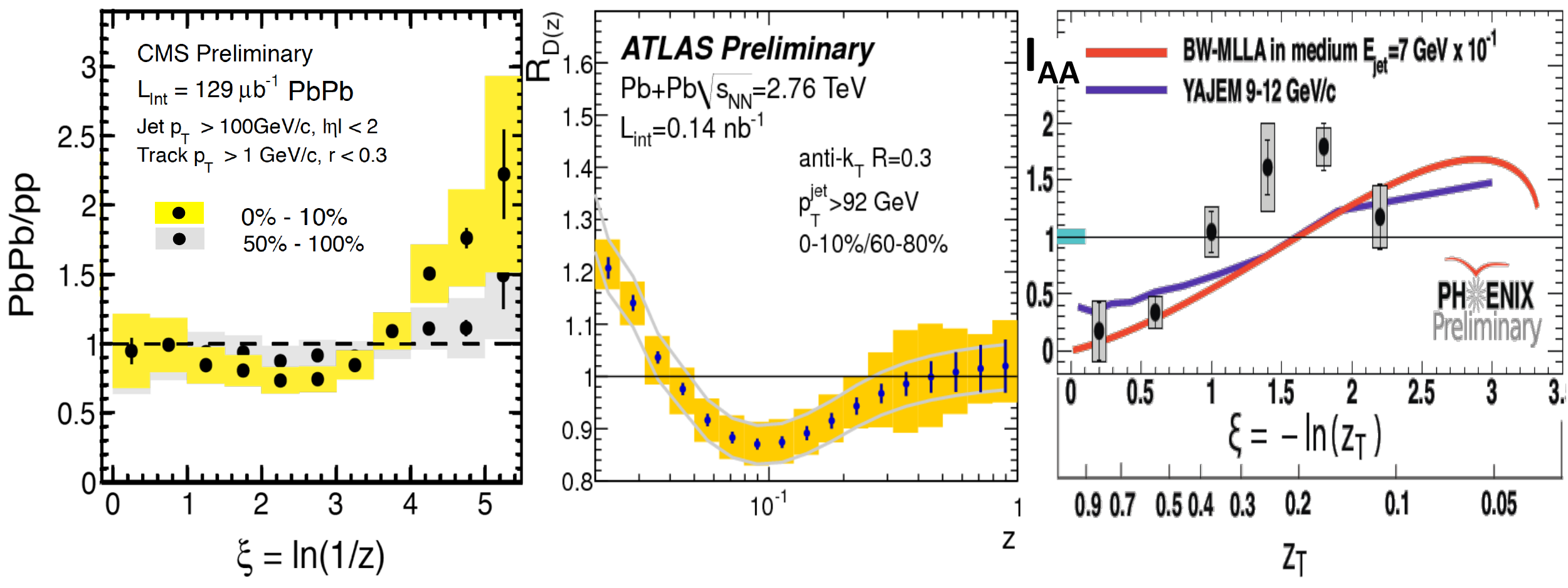}
\end{center}
\vup
\caption{Left: ratios of fragmentation functions measured by the CMS experiment in two centrality bins to the same in \pp~\cite{cms-frag}. Middle: central-to-peripheral ratio measured by ATLAS~\cite{atlas-frag}. Right: the $I_{AA}$ from PHENIX~\cite{phenix-frag}. Data are compared to the computations~\cite{Renk:2011wp,Borghini:2005em}.}
\label{fig:jets_ff}
\end{figure}

%%%%%%%%%%%%%%%%%%%%%%
Among the most interesting results presented at the conference are the first high statistical measurements of the jet fragmentation function~\cite{cms-frag,atlas-frag,phenix-frag} shown in the three panels of Fig.~\ref{fig:jets_ff} versus $z=\vec{\pt}^{charged}\cdot\vec{\pt}^{jet}/|\vec{\pt}^{jet}|$ or versus $\xi=-ln(z)$.
The results presented by the LHC experiments use the reconstructed $\pt^{jet}$ which includes the effect of jet energy loss. The $y$-axis in the two left panels is the ratio of fragmentation functions measured in \pbpb\ centrality bins to \pp\ for CMS and to peripheral centrality bin for ATLAS. In spite of the different representation, the results agree. For $z\rightarrow1$ ($\xi\rightarrow0$) the modification of the fragmentation function is small, and the ratio is consistent with unity. The fragmentation function ratio reaches its minimum just below $z\approx0.1$ ($\xi\approx3$) and grows significantly above unity towards small $z$.
PHENIX result~\cite{phenix-frag} shown in the right panel uses the ratio of charged particle momentum to the \pt\ of a photon going in the opposite direction as an estimate of  the jet energy before energy loss. The ratio of the per-trigger yield in \auau\ and \pp\ collisions ($I_{AA}$) is significantly lower than unity at $\xi\rightarrow0$ and  grows with $\xi$.

There is a difference in determination of the fragmentation functions shown in the left two panels of Fig.~\ref{fig:jets_ff} and theoretical predictions, such as~\cite{renkproc} presented at  this conference, which use jet \pt\ prior to energy loss to define $z$. This prevents a direct comparison of theory and the data, however results in Fig.~\ref{fig:jets_ff} support the  generic expectation of a softening of the fragmentation process. Nevertheless,  one must  keep in mind that due to absolute normalisation of the fragmentation function, one should expect a correlation in different $z$ regions which may complicate this conclusion.
Furthermore,  the fact that these functions remain unchanged at high $z$ with respect to the \pp\ baseline seems puzzling: while it is clear that  the presence of a high-$z$ particles dominate the jet energy it is not clear why the expected softening does not alter this high-$z$ distribution; input from available models is needed before drawing a conclusion on this subject. Contrary to these observables, since the photon \pt\ is a better estimator of the jet momenta prior to energy loss, a direct comparison of the photon hadron correlation measurement in Fig.~\ref{fig:jets_ff}  with the theoretical calculations in~\cite{renkproc, Renk:2011wp,Borghini:2005em} can be made, which verify once again the generic expectations of an enhancement of the soft components of the fragmentation function.  Note that while the two model calculations shown in the figure are based on a modification of the in-medium splittings, the implementation of such modification is very different. Those implementations reflects the theoretical uncertainties in the treatment of this process, which may partially explain the quantitative disagreement between the data and the curves shown in Fig.~\ref{fig:jets_ff}. 

In addition,  it has also been pointed out that in both weak and strong coupling plasma models, soft particles in the medium become correlated with the jet direction as a consequence of the jet-medium interaction~\cite{colproc,krishnaproc}. This fact  complicates the interpretation of in-medium fragmentation functions as arising from radiative QCD processes, specially for the soft jet components.
\\

%%%%%%%%%%%%%%%%%%%%%%%%%%%%%%%%%%%%%%
Quark Matter 2012 provided a wealth of new experimental results on high-\pt\ observables in HI collisions including the first precision measurements of the jet fragmentation, boson-jet correlations, angular anisotropy of high-\pt\ charged particles and jets, the first results on the suppression of b-jets. Further studies would greatly benefit from larger statistics, which in many cases remains the limiting factor. Understanding the underlying physics requires the extension of the kinematic coverage, especially at low-\pt\ and detailed comparison between the LHC and RHIC energy regimes.

Several models successfully describing different aspects of experimental data were discussed at the conference. The differences between the models primarily relate to the implementation of the shower-medium interaction. One of the most salient distinctions is the treatment of the evolution time structure. Some models like~\cite{colproc} assume that all shower fragments arise prior to interactions with the medium. Other models~\cite{vitevproc,zangproc} consider the fragmentation to happen outside of the medium as in vacuum. Some models, e.g.~\cite{renkproc}, use a formation time argument to decide whether  splittings  occur inside or outside of the medium. These differences arise from a lack of theoretical control over the time structure of the vacuum fragmentation process which must be addressed beyond parametric estimates. Most of current models do not include interference effects between multiple color charges  originating from the fragmentation process   and propagating inside the medium.  Effects of the medium   on the interference pattern between emissions of those charges control the ability of the medium to resolve more than one color source propagating in the system.  Very promising studies in this direction show a non-trivial interference pattern that depends on the transverse separation of the pair in the medium ~\cite{yacineproc}. These effects, which have not yet been introduced into models,  reveal that there may be many interesting phenomena that arise as a consequence of the multipartonic nature of jets.
%{\magenta
\\
%we need to fill this line 0\\
%we need to fill this line 1\\
%we need to fill this line 2\\
%we need to fill this line 3\\
%we need to fill this line 4\\
%we need to fill this line 5\\
%we need to fill this line 6\\
%}

The work of A.M. is supported by FP7-PEOPLE-IRG (grant 710398), Minerva Foundation (grant 7105690) and by the Israel Science Foundation (grant 710743). The work of JCS is supported by a Ram\'on y Cajal fellowship and  by the research grants FPA2010-20807, 2009SGR502 and by the Consolider CPAN project.

\end{document}